\documentclass[referee,sn-basic]{sn-jnl}

\usepackage[utf8]{inputenc} 
\usepackage[T1]{fontenc}    
\usepackage{hyperref}       
\usepackage{url}            
\usepackage{booktabs}       
\usepackage{amsfonts}       
\usepackage{nicefrac}       
\usepackage{microtype}      
\usepackage{lipsum}
\usepackage{amsmath}
\usepackage{fancyhdr}       
\usepackage{graphicx}       
\usepackage{natbib}
\usepackage{setspace}
\usepackage{xcolor}
\graphicspath{{media/}}     

\usepackage{xr}
\usepackage{cleveref}

\usepackage[ruled,vlined]{algorithm2e}
\makeatletter
\renewcommand{\algocf@caption@ruled}{%
  \box\algocf@capbox\par\vskip\interspacetitleruled%
  \hrule height1.2pt depth0pt \kern\interspacealgoruled 
}
\makeatother

\begin{document}

\title[Article Title]{Combining opinion and structural similarity in link recommendations to counter extreme polarization}


\author*[1]{\fnm{Gabriella} \sur{Dantas Franco}}\email{g.dantasfranco@uva.nl}

\author[1]{\fnm{Marta} \sur{C. Couto}}\email{m.gomesdacunhacouto@uva.nl}

\author[1]{\fnm{Vítor V.} \sur{Vasconcelos}}\email{v.v.vasconcelos@uva.nl}

\author[1]{\fnm{Fernando} \sur{P.  Santos}}\email{f.p.santos@uva.nl}

\affil[1]{\orgdiv{Informatics Institute}, \orgname{University of Amsterdam}, \orgaddress{\city{Amsterdam}, \postcode{1098 XH}, \country{Netherlands}}}



\abstract{Recommendation algorithms, used in online social networks, shape interactions between users. In particular, link-recommendation algorithms suggest new connections and affect how individuals interact and exchange information. 
These algorithms' efficacy relies on key mechanisms governing the creation of social ties, such as triadic closure and homophily. The first is achieved through \emph{structural} similarity and represents a heightened chance of recommending users to one another given mutual friends; the second is related to \emph{opinion} similarity and conveys an increased chance of recommending a connection given similar individual characteristics. 
These two mechanisms jointly shape the evolution of social networks and behaviors unfolding over them. Their combined effect on the co-evolution of opinion and structure dynamics remains, however, poorly understood.
Here, we study how social networks and opinions co-evolve given the joint effect of rewiring based on opinion and structural similarity. We show that both similarity metrics lead to polarized states, but differ in how they impact network fragmentation and opinion diversity. While strongly relying on opinion similarity leads to a higher variation of opinion, rewiring via network similarity leads to a larger number of (dis)connected components, resulting in fragmented networks that lean towards one of the signed opinions.
Under strong homophilic settings, introducing a weak dependence on structural similarity prevents network fragmentation and favors moderate opinions.
This work can inform the design of new recommender algorithms that explicitly account for interacting social and recommendation mechanisms, with the potential to foster moderate opinion coexistence even in inherently polarizing settings.}

\keywords{Complex systems, recommender algorithm, co-evolving network and opinions}

\maketitle

\section{Introduction}

Online social networks (OSNs) mediate a large fraction of our social and professional interactions, with various implications for people's offline lives. OSNs affect  political mobilization \citep{bond201261, jones2017social},  health and well-being \citep{hobbs2016online, burke2016relationship}, or even access to social and financial capital \citep{chetty2022socialI, chetty2022socialII, steinert2015online, greijdanus2020psychology, rajkumar2022causal}. In online environments, new connections are increasingly shaped by algorithmic recommendations. In particular, link recommendation algorithms are used to recommend new connections to users, influencing who connects with whom, how information diffuses, and ultimately how the structure of OSNs evolves \citep{daly2010network,su2016effect,su2020experimental,santos2021link}. 

In order to provide recommendations that users accept, link recommendation algorithms rely, explicitly or implicitly, on well-known mechanisms for social ties formation.  \emph{Homophily} and \emph{triadic closure} are two key examples. These mechanisms have long been studied in network science as key principles for the formation of social bonds \citep{louch2000personal}. Similarly, homophily and triadic closure suggest important heuristics in link-prediction algorithms \citep{liben2003link,li2017survey} and are implicitly leveraged by more complex graph representation learning methods \citep{grover2016node2vec,kipf2016variational}. Public information offered by large platforms to explain algorithmic link recommendations commonly identifies shared connections and common interest as key metrics used, which correspond closely to triadic closure and homophily in network-science terms \citep{Meta2026transparency,Twitter2026transparency,LinkedIn2026transparency}. Despite capturing different aspects of social connections and being jointly used in link recommendations, the combined effects of homophily and triadic closure remain poorly understood.

Homophily is an umbrella term used to describe the tendency of individuals to interact with similar others \citep{lazarsfeld1954friendship, mcpherson2001birds}. When designing link-prediction or link-recommendation methods, homophily implies that links should be suggested based on \emph{opinion similarity}, i.e., a similarity function comparing nodes based on their opinions, preferences, or attitudes. This metric operates on individuals' properties, accounting for user characteristics that are inferred from their online interactions \citep{aiello2012friendship, sajjadi2024unveiling}. Triadic closure, on the other hand, assumes that two disconnected individuals are likely to become linked if they share connections  \citep{peixoto2022disentangling,bachmann2025network}. In general, triadic closure is based on a similarity function that assumes that two nodes are similar if they share a high number of neighbors. Such similarity function is often called \emph{structural similarity} as it only depends on the social network structure \citep{lu2011link} -- in opposition to opinion similarity, which depends on individuals' preferences and opinions.  

Understanding how rewiring based on opinion and structural similarity shape network structure and opinion dissent is therefore essential for assessing the long-term effects of link recommendation systems. 
Disentangling homophily and triadic closure as causal mechanisms in empirical data has only been addressed recently by \cite{peixoto2022disentangling}. Moreover, their cumulative effects have been investigated by \cite{asikainen2020cumulative} on a population with static opinions. More theoretically, \cite{abebe2022effect} have studied the effect on network segregation when adding an edge via triadic closure on stochastic block models with homophily. \cite{bachmann2025network} studied the joint effects of triadic closure and homophily in inequality between groups. In all previous works mentioned, homophily depends on individual characteristics and group membership that remain static over time. In another direction, previous work studied coevolution of node properties and network topology \citep{gross2008adaptive} and it has been shown that, when opinions co-evolve with homophilic rewiring, there is a continuous phase transition in structure \cite{holme2006nonequilibrium}. However, we still lack understanding of the joint effects of homophily and triadic closure in recommender systems, particularly in domains where both the network and opinions -- or, broadly, nodes' attributes -- co-evolve. We are compelled to understand the co-evolution of node attributes and network structure as this setup can capture ubiquitous phenomena in OSNs such as feedback loops, the emergence of echo-chambers and the emergence of polarized social structures detrimental to collective action \citep{vasconcelos2021segregation}. Disentangling peer influence effects from homophily and triadic closure, all of which lead to dense networks of like-minded individuals, is particularly hard empirically due to confounded mechanisms and various feedback loops \citep{su2020experimental}.

To address this gap, we propose a model in which opinions and network structure co-evolve. Opinions are updated based on neighbors average influence following an update rule recently proposed as capturing the formation of echo chambers in online social media platforms \citep{baumann2020modeling}. Link rewiring, on the other hand, is driven by a tunable combination of opinion and structural similarity. By tuning both the strength and interpolation between opinion and structural similarity, we aim at testing mechanistically different forms of link-recommendation applied in OSNs. The co-evolving dynamics of opinions and links leads to distinct regimes of polarization and network fragmentation, highlighting the importance of algorithmic design choices. We ask whether algorithms that prioritize structural similarity over opinion similarity lead to qualitatively different patterns of network connectivity and whether certain parameter regimes sustain opinion coexistence without extreme polarization and network segregation.

Our results reveal differences between structural and opinion-driven rewiring. Both lead to polarized and fragmented communities that hold opposing opinions and are not connected to each other. However, opinion-based rewiring leads to a fragmented state more easily, with a lower threshold for the fragmentation transition. We also show that a weak reliance on structural-based triadic closure can preserve a connected network, allowing for the coexistence of moderate opinions within the same interacting structure. This highlights that a small amount of structural similarity might be needed to counteract the inherently polarizing aspect of homophily.

\section{Methods}
\label{sec:Methods}

To investigate the effects of link-recommendation algorithms on population-level opinion and network structure, we develop a mechanistic co-evolutionary model that accounts for rewiring based on opinion and structural similarity metrics. We aim to understand their impact on long-term outcomes in the user population. In this section, we detail the models used in this study. First, we present the recommender design, which interpolates between opinion and structural similarity. Secondly, we introduce the opinion dynamics model. Lastly, we discuss the metrics used to capture opinion dynamics.

\subsection{Network dynamics through link-recommendations}\label{subsec::recommender}

Link-recommendation algorithms used in OSNs involve complex algorithms and pipelines \citep{Meta2026transparency,LinkedIn2026transparency,twitter_follow_recommendations_service,gupta2013wtf}. Fundamentally, however, these systems often rely on a combination of users' activity, interests and network data. Here, we attempt to isolate, in a stylized manner, two key metrics used in link-recommendations (explicitly as heuristics or implicitly in complex graph learning methods): opinion similarity and structural similarity. We assume that nodes will be ranked, and and top candidates are offered as recommendations to users. The selection process is simulated by a random draw from the probability distribution of recommended not-yet-friends, and we assume that, once drawn, the recommended user is accepted as a friend, establishing the new connection. Once a new link is formed, the focal individual will choose one of its previously existing connections to break, with uniform probability. This assumption is based on the `attention-budget phenomena', an expression commonly used to state that despite the increasing number of connections on OSNs, each individual is only able to reciprocate attention with a few other users \citep{huberman2008social, weng2012competition}. It also keeps the number of total edges constant over time, which facilitates analytic approaches. 

The population is represented by an undirected graph $G$, where each node $u$ is a user ($u \in V$), and each link $l$ is an active connection ($l \in E$). $E$ denotes the set of edges (links) and $V$ the set of vertices (nodes/users). The network changes are captured by a dynamic adjacency matrix $A_{ij}^t$, where $A_{ij}^t=1$ if an edge between node $i$ and node $j$ exists at time $t$, and $A_{ij}^t=0$ otherwise.  
To interpolate between recommendations based on opinion similarity and structural similarity, we assume that the probability of adding a new link between nodes $i$ and $j$ is given by 
\begin{equation}
    P[A_{ij}^{t+1}=1] = \rho\, H_{ij}(t) + (1-\rho)\, S_{ij}(t),
\end{equation}
where $H_{ij}$ and $S_{ij}$ represent the similarity measures based on opinion and structure, respectively. The parameter $\rho$ can be interpreted as the weight of opinion similarity on recommendations, in opposition to structural similarity. At one extreme, $\rho=0$ represents a rewiring dynamics that is purely guided by the structural similarity metric, recovering \citep{santos2021link}. On the other extreme, $\rho=1$ considers only opinion similarity, and does not take structure into account, approximating models where homophily shapes the rewiring mechanism \citep{holme2006nonequilibrium,blex2022positive,borges2024social,vasconcelos2019consensus,gross2008adaptive}. 

At each time step, each node $i$ will have a turn as focal, selecting a new node $j$ to add to its neighborhood set $N_i$. Once selected, the friendship is accepted with probability $1$, leading to $A_{ij}^{t+1}=1$. The new friendship connection is drawn from the focal node's recommendations, given by the probability distribution $P[A_{ij}^{t+1}=1]$ defined above.

\subsubsection{Structural similarity}
For the purpose of this study, structural similarity between nodes $i$ and $j$ is represented by the number of common neighbors, or the cardinality of the intersection between the neighborhood sets of $i$ and $j$, $|N_i \cap N_j|$. 
Although simple, this metric has been used extensively in link-prediction methods with great success \citep{liben2003link}. Common neighbors statistics have been estimated from empirical network data in frameworks such as Stochastic Actor-Oriented Models \citep{snijders2001statistical, goodreau2009birds}, lending plausibility to its use as an empirically testable mechanism. 
Moreover, to allow for links between different groups to be formed, we include a fixed amount of noise $\epsilon$, that can also be interpreted as algorithmic uncertainty. A non-zero noise is also needed to avoid divergences on the similarity metrics. For all nodes $j$ not currently connected to the focal $i$, we then have that
\begin{align*}
    S_{ij}(t) &= \frac{\big[|N_i^t\cap N_j^t|(1-2\epsilon)+\epsilon \big]^{\eta}}
    {\sum_{k\notin N_i}\big[|N_i^t\cap N_k^t|(1-2\epsilon)+\epsilon \big]^{\eta}},
\end{align*}
where $\eta(\geq0)$ controls the strength of the similarity metric. The higher this exponent, the more skewed the distribution is. 

\subsubsection{Opinion similarity}
We construct the opinion similarity metric similarly, but taking into consideration the opinion space instead of the network structure. As detailed below, we assume that 
each individual has a real-valued opinion $x_i\in \mathbb{R}$ over a single, isolated topic.
Given this one-dimensional opinion space, a natural proxy for homophily is the distance between individual opinions -- subject to the same noise $\epsilon$. Analogously to the structural similarity function, there is a non-linear parameter $\beta(\geq0)$ representing the strength of similarity, leading to
\begin{align*}
    H_{ij}(t) &=\frac{\big[|x_i^t-x_j^t|(1-2\epsilon)+\epsilon \big]^{-\beta}}
    {\sum_{k\notin N_i}\big[|x_i^t-x_k^t|(1-2\epsilon)+\epsilon \big]^{-\beta}}, 
\end{align*}
where now $|x|$ denotes the absolute value of $x$.

Once again, the metric is normalized for each focal node $i$, so that each user selects exactly one candidate from its distribution. The negative sign on the exponent turns the distance between opinions, which measures how dissimilar the node opinions are, into a similarity metric.\\

This setup fully describes the recommender with only a few parameters: the exponents $\eta$ and $\beta$ independently tune the strengths of structural and opinion similarity. Importantly, these parameters allow us to test different dependencies of recommendations on similarity metrics: a sub-linear ($0<\eta<1$), linear ($\beta,\eta=1$), and supra-linear ($\beta,\eta>1$) dependency, later shown to be important. Additionally, $\rho \in [0,1]$ balances the weight of opinion similarity against structural similarity. And $\epsilon$ defines algorithmic noise, which is considered constant. For some more discussion over $\epsilon$ and its effect on this dynamics, please refer to the Supplementary Information.

\begin{figure}
    \centering
    \includegraphics[width=0.95\linewidth]{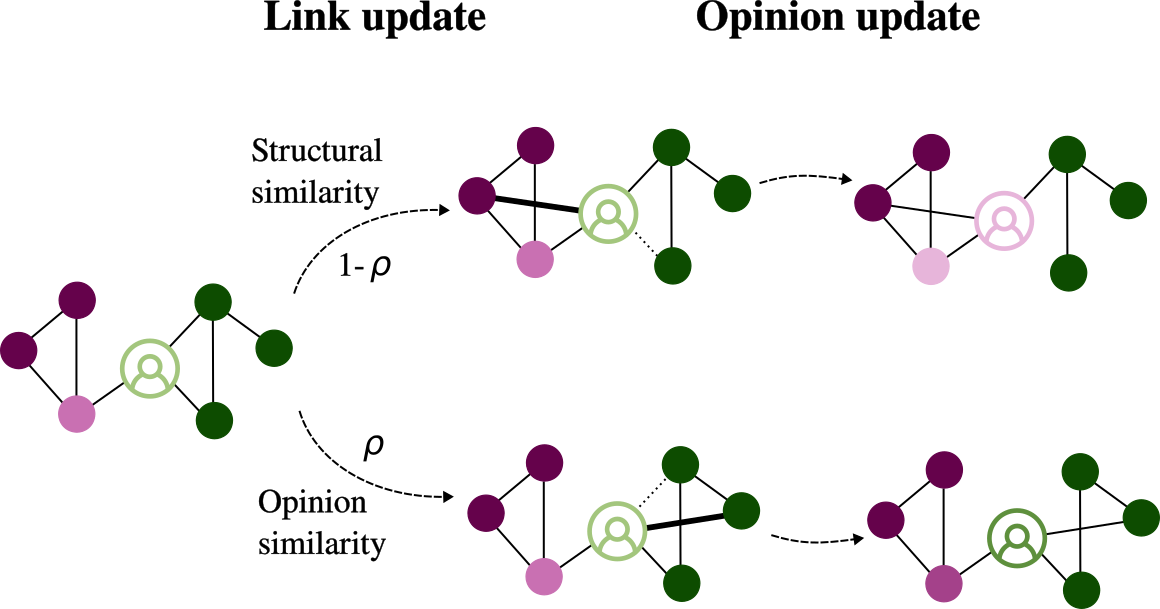}
    \caption{Illustration of our model. In this example, we assume that the center node (light green) will rewire one of its connections. This node receives a friend recommendation based on opinion similarity ($\rho$) or structural similarity ($1-\rho$) and accepts it (thick edge). At the same time, it chooses a random existing link to exclude (dotted edge). Once the whole network has had a chance to rewire, all nodes update their opinion simultaneously, as indicated by the color changes, where darker colors indicate more extreme opinions.}
    \label{fig:diagram}
\end{figure}

\subsection{Opinion Dynamics}

As introduced above, we aim at understanding the effects of link recommendations on a population with evolving opinions. We adopt the discrete-time counterpart of a model recently identified as suitable to model echo chambers in online social networks \citep{baumann2020modeling}. This model assumes that opinions change given the average opinion on a node's neighborhood; under random rewiring, this type of local averaging dynamics leads to consensus being an absorbing state \citep{degroot1974reaching}.

Each subject $i$ has an individual opinion $x_i^t\in \mathbb{R}$ about a single topic. 
Social influence is modeled as a synchronous process: once every user has had the opportunity to rewire one of its connections, all of them revisit their opinions at the same time, influenced by their respective neighborhoods. As introduced above, we consider the average opinion of a node's neighborhood; this effect is however non-linear: users with extreme opinions have a bounded amount of influence over their peers. This assumption is supported by empirical evidence \citep{jayles2017social} and has been studied in the context of online activity-driven networks \citep{baumann2020modeling, baumann2021emergence}. 

In the absence of peer influence, individuals forget the topic, and their opinion decays towards a neutral stance at a constant rate $0\le\gamma<1$. This can be interpreted as referring to a topic that is not a core characteristic of the individuals: it is only important when others are talking about it, much like `hot topics'. 

Building on these observations, formally, the opinion dynamics we consider is given by
\begin{equation}
    x_i^{t+1} = \gamma\, x_i^t + K \sum_{j=1}^N \frac{A_{ij}^t}{k_i^t}\tanh(\alpha\, x_j^t),
\label{eq:OD}
\end{equation}
where $k_i^t = \sum_l A_{il}^t$ is the degree of node $i$ and $K$ controls social influence. Lastly, $\alpha$ controls how one's opinion translates into social influence, representing the interplay between issue controversy and social reinforcement of this model \citep{baumann2020modeling, santos2021link}. We illustrate the whole model in Fig. \ref{fig:diagram}.

On a static network (or under random rewiring), low $\alpha$ leads to a stable solution of neutral consensus ($x_i=0,\ \forall i$); for higher values, persistent non-neutral stances emerge \citep{gray2018multiagent, bizyaeva2022nonlinear}.  
Although we allow any real value for the opinion variable, the balance between opinion decay and social influence effectively bounds the opinions in the range $[\frac{-K}{1-\gamma},\frac{K}{1-\gamma}]$.

Unless otherwise stated, we set the parameters $K=0.1$, $\gamma=0.99$ and $\alpha=0.3$ (see \citep{baumann2020modeling, baumann2021emergence,santos2021link} for more details on this model's behavior and interpretation on online social settings). Under these values, any connected static or randomly rewired network converges to a non-neutral but consensual state. When connections adapt through the recommender system, however, this long-term outcome may change -- so that \emph{any polarization observed, either in opinion or in network structure (e.g., fragmentation), arises solely from the network dynamics and, to a large extent, the link-recommendation algorithm assumed.}

\subsection{Metrics}\label{subsec::metrics}

We use three main metrics to interpret our results: polarization, radicalization and the number of connected components.
First, we are interested in characterizing a real-valued stance within a population, so it is sensible to choose a dispersion (or variance-)based measurement for the opinions. For that purpose, we define \textbf{polarization} as
\begin{equation*}
    \text{Polarization}(\mathbf{X}) \equiv \text{std}(\mathbf{X}),
\end{equation*}
i.e., the standard deviation of the opinion vector $\mathbf{X}=(x_1, x_2, \dots, x_N)$. 
As a dispersion-based measure, this definition of polarization is a good representation of dissent in a population. Polarization is only null when there is complete consensus.

We are also interested in how extreme or radical the opinions can be.
Previous work considers the mean of absolute opinions as the radicalization level within a population \citep{santos2021link}. We follow a similar approach and define \textbf{radicalization} as
\begin{equation*}
    \text{R}(\mathbf{X}) \equiv \frac{1}{N}\sum_{i=1}^{N}|x_i|.
\end{equation*}

Polarization is a variance-based metric that can take into account both in-group dispersion and between-group dispersion.
However, many distinct network structures can display similar opinion distributions.
As such, this metric does not capture \emph{structural polarization}, as described in  \citep{borges2024social}. There, besides attribute-level dissent, there is also a network-level separation, or fragmentation. 
Therefore, to account for fragmentation, we also consider the \textbf{number of connected components} of the network. A connected component of $G$ is a maximal connected subset of vertices $C \subseteq V$, i.e., for all $u,v \in C$ there exists a path between $u$ and $v$, and no vertex in $V \setminus C$
is connected by a path to any vertex of $C$.

\section{Results}
\label{sec:results}

\begin{figure}[ht]
    \centering
    \includegraphics[width=0.65\textwidth]{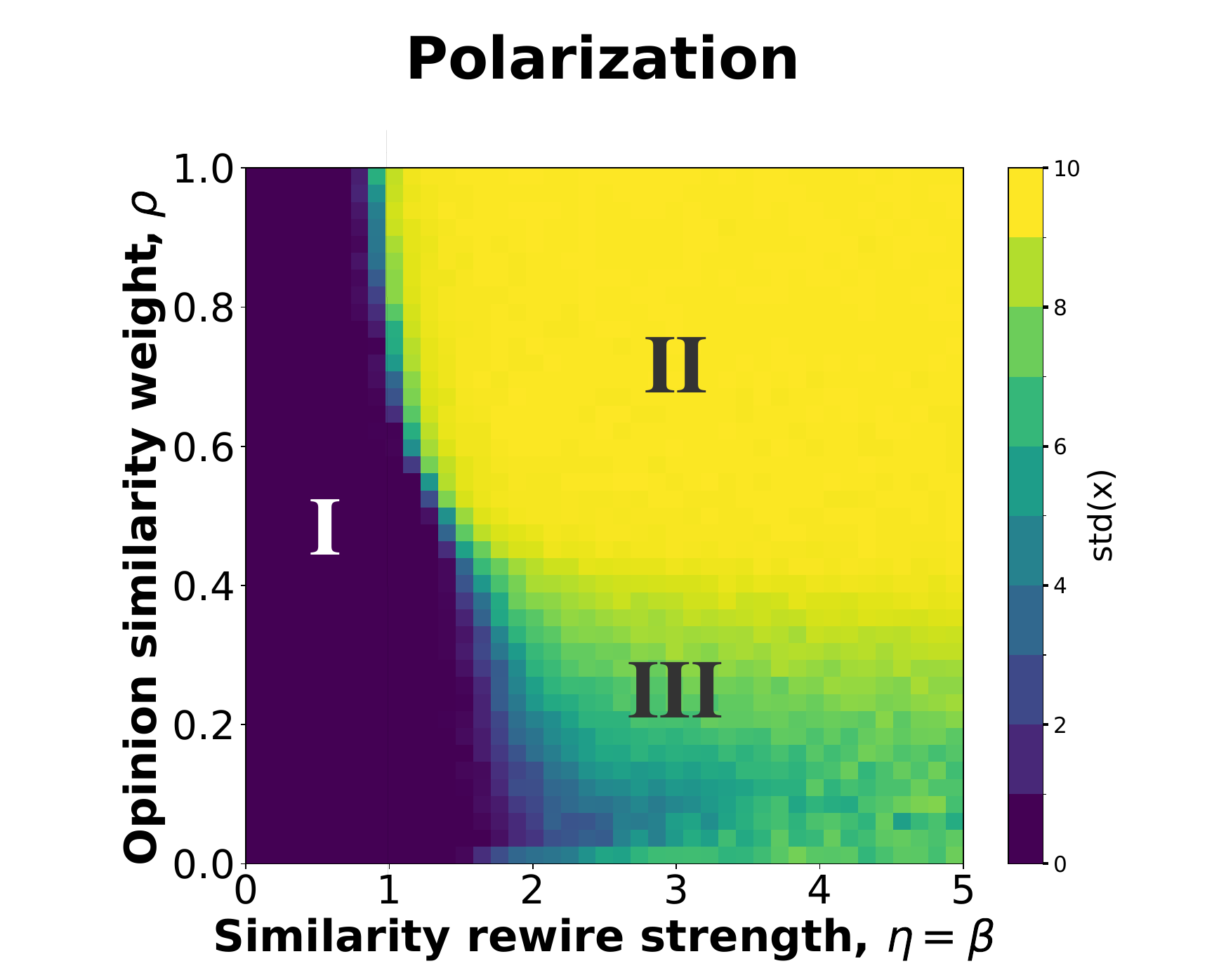}
    
    \includegraphics[width=1.0\textwidth]{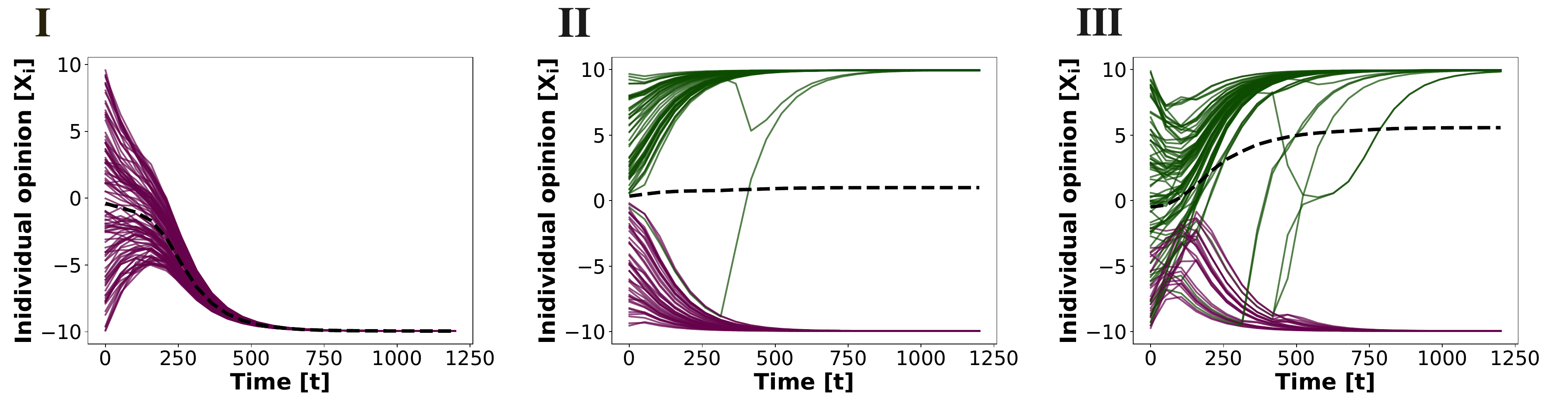}
    \caption{(Top) Polarization when varying the dependency of recommendations on opinion and structural similarity (x-axis, $\eta=\beta$) and the relative importance of opinion similarity (homophily) on recommendations (y-axis, $\rho$). Each point corresponds to the average polarization over 50 independent simulations. We observe that polarization emerges for any value of $\rho$, given a sufficiently high value of $\beta$ and $\eta$ (here we assume $\beta = \eta$). This suggests that strong homophily and triadic closure each independently contribute to opinion polarization.  
    (Bottom) Timeseries of opinion evolution for three paradigmatic regimes highlighted in the top panel: I ($\eta=0.5$, $\rho=0.5$), II ($\eta=3.0$, $\rho=0.75$), III ($\eta=3.0$, $\rho=0.25$). Each line represents an agent's opinion over time, colored according to its value at $t=1200$. Dashed lines indicate the average opinion of the population. We confirm that strongly relying on similarity-based rewiring results in a bi-modal and highly polarized opinion landscape. Other parameters considered: $\alpha=0.3$, $K=0.1$, $\gamma=0.99$, $t_{max}=1200$.}
    \label{fig:Pol_rhoeta}
\end{figure}

Recall that the effect of opinion and structural similarity-based metrics on recommendations is controlled by two parameters ($\beta$ and $\eta$, respectively). Furthermore, the relative importance of opinion similarity over structural similarity is controlled by $\rho$. 
Initially, we assume that $\eta=\beta$, allowing us to analyze the emerging levels of polarization for the 2-dimensional parameter space ($\eta=\beta,\ \rho$) shown in Fig. \ref{fig:Pol_rhoeta}. First, we note that, for any value of $\rho$ -- that is, whether the re-linking prioritizes the opinion or structural similarity -- there exists a threshold on similarity strength ($\beta$, $\eta$) after which polarization emerges. This suggests that both similarity mechanisms
can contribute to polarization. However, the highest values of polarization ensue when recommendations prioritize opinion similarity (high $\rho$), a setting where we observe a sudden transition between consensus and polarization.

The bottom panels of Fig. \ref{fig:Pol_rhoeta} illustrate typical opinion timeseries for the highlighted parameter regions. In region I (leftmost), below the critical value for the similarity strength, opinion consensus emerges. Under this parameter regime, any random uniform initial condition converges with equal chance to the extreme negative or positive stance, similarly to random recommendations. In region II (upper-right), which corresponds to opinion similarity being favored in recommendations ($\rho>0.5$), the average opinion over the population remains close to null. The transition from consensus to polarized states occurs near the linear recommendation ($\eta=\beta=1$). On the other hand, region III (lower-right) showcases how prioritizing structural similarity in recommendations leads to the average opinion over the population being biased towards one direction -- explaining partly why we observe lower polarization when $\rho$ is low. Here, the threshold values of $\eta$ above which polarization is non-zero are supra-linear. This suggests that consensus persists for a wider range of similarity strengths when recommendations favor structural similarity.\\ 

\begin{figure}[t]
\centering
    \includegraphics[width=0.70\linewidth]{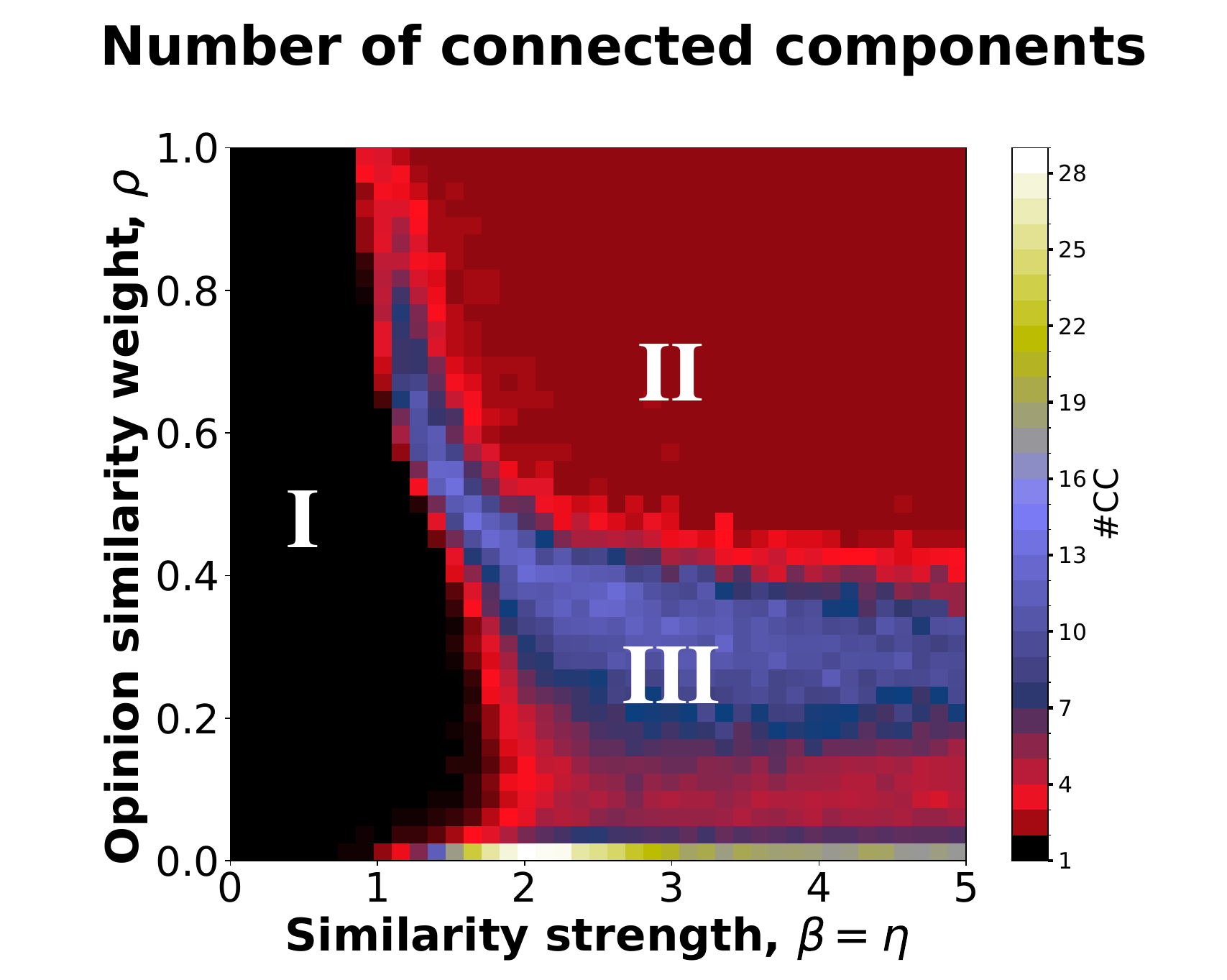}
    
    \includegraphics[width=0.25\linewidth]{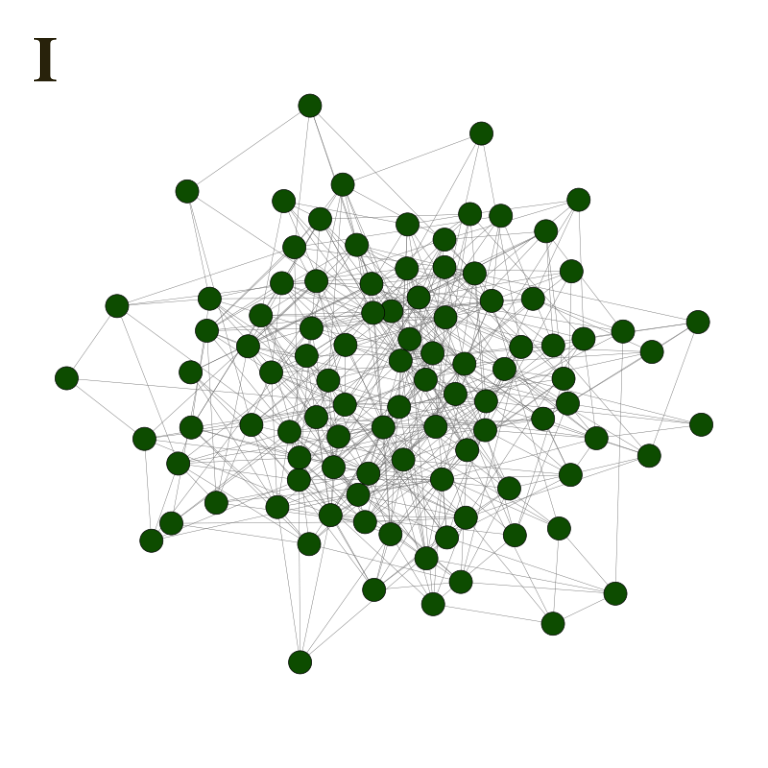}
    ~
    \includegraphics[width=0.25\linewidth]{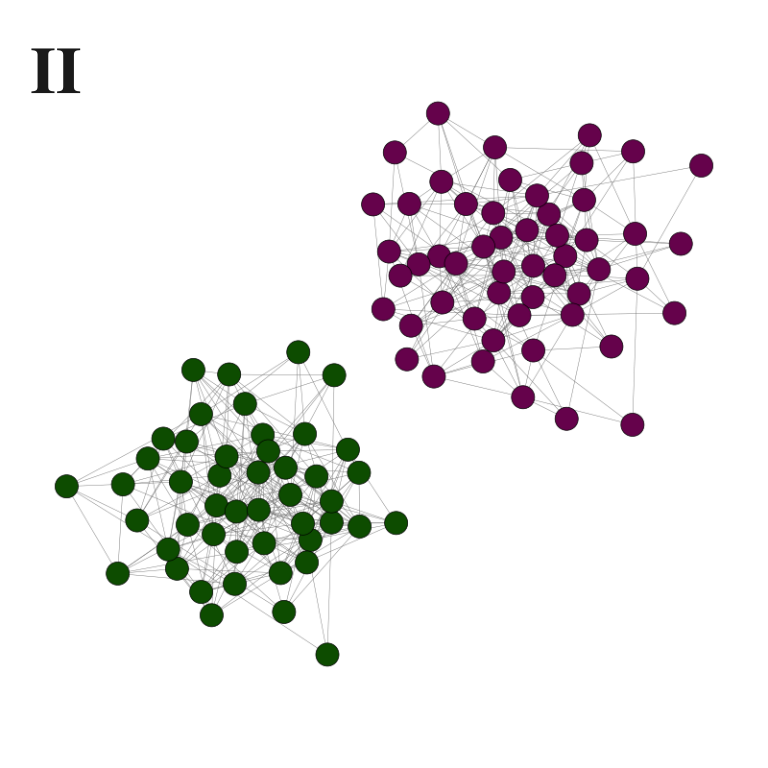}
    ~
    \includegraphics[width=0.25\linewidth]{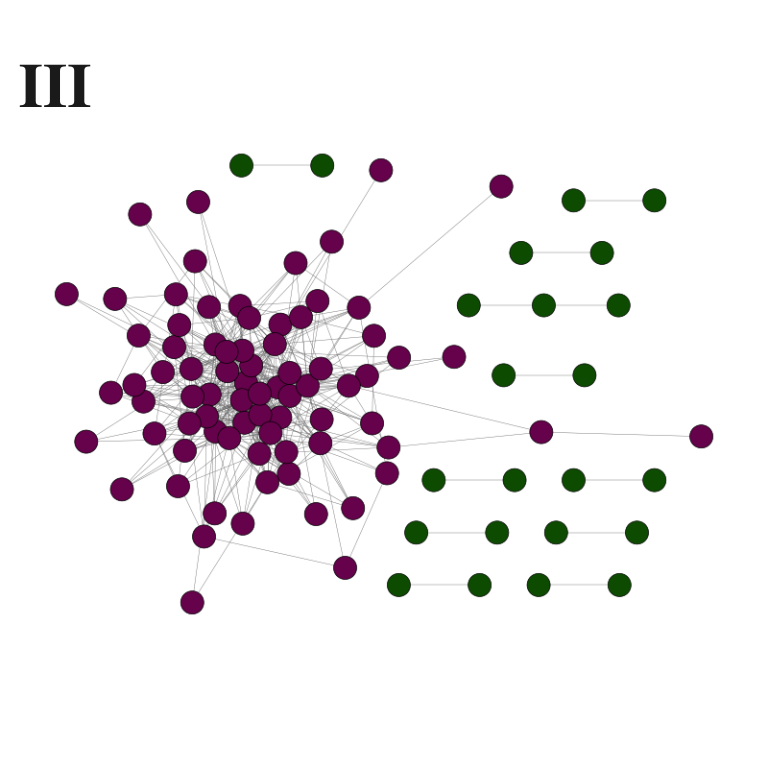}
    ~
    \includegraphics[width=0.12\linewidth]{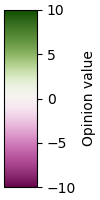}
    \caption{Strong homophily and transitivity each independently contribute to opinion polarization, however this occurs through the emergence of different network topologies. (Top) Average number of connected components ($\#CC$) over $50$ independent simulations. The color scale emphasizes the regions with: (I) a single connected component; (II) approximately two connected components, which corresponds to prioritizing homophily ($\rho>0.5$); and (III) many connected components, which appear when the recommender prioritizes transitivity ($\rho<0.5$), showing a high level of fragmentation. (Bottom) Long-term network structures for the three paradigmatic regimes highlighted in the parameter space. Other parameters considered: $\alpha=0.3$, $K=0.1$, $\gamma=0.99$, $t_{max}=3000$. }
    \label{fig:NCC_RhoEta}
\end{figure}

Inspecting polarization results alone might suggest that opinion and structural similarity play qualitatively similar roles in leading to polarized communities, apart from differences in polarization values and response to changes in similarity rewiring strength. However, looking at the number of connected components over the same parameter space, shown in Fig. \ref{fig:NCC_RhoEta}, we see that fragmentation levels differ greatly depending on the recommender's dependency on opinion or structural similarity. For $\rho<0.5$, the number of connected components is high, having two peaks: a global maximum at $\rho=0$ and a local one at intermediate values of $\rho$. 
For $\rho>0.5$, the system generally reaches a stable structure of $2$ connected components that do not exchange information, which is expected in highly homophilic systems (as investigated by \cite{baumann2021emergence}, for example). We also note that such fragmented regime is reached with a lower similarity strength when the recommender prioritizes opinion similarity. This emphasizes that, in the case of equal similarity strengths ($\beta=\eta$), the network is more resilient to fragmentation when recommendations rely more on structural similarity. 

Examples of the resulting networks in each region are displayed on the bottom panels of Fig. \ref{fig:NCC_RhoEta}. These network examples depict two general trends observed: 
1) High similarity strength with the recommender relying on opinion similarity leads to two groups of roughly the same size;
and 
2) High similarity strength with the recommender relying on structural similarity leads to a fragmented state with a tightly connected majority group, and many small opposing communities.

The previous results indicate that when both $\eta$ and $\beta$ are high (i.e., recommendations are highly sensitive on opinion and structural differences between users) interpolating between these two features (changing $\rho$) leads to significantly different network topologies. This means that, even if grounding link recommendations on either opinion or structural similarity can increase polarization, the distribution and magnitude of opinions can vary as a result of different network configurations. Understanding how to combine high values of $\eta$ and $\beta$ to mitigate the negative effects of polarization is particularly relevant in our context as one needs to guarantee that recommendations are still accurate and useful for users \citep{li2017survey}, thereby relying on their interests and acquaintances (i.e., $\beta=\eta=0$ might lead low polarization, however such recommenders will hardly be adopted given their random nature). 

Beyond polarization, we now turn our attention to radicalization, a measure of how extreme opinions are. We start by noting that polarization and radicalization can vary independently: even if radicalization is high, polarization depends on opinions' dispersion within opposing signed groups, on the relative group sizes and the exact magnitude of opinion within each group (we formally describe the relationship between polarization and radicalization in the Supplementary Information, Appendix A). We thereby ask: how to set $\rho$ under high values of $\beta$ in order to reduce radicalization?

\begin{figure}[t]
    \centering
    \includegraphics[width=0.9\textwidth]{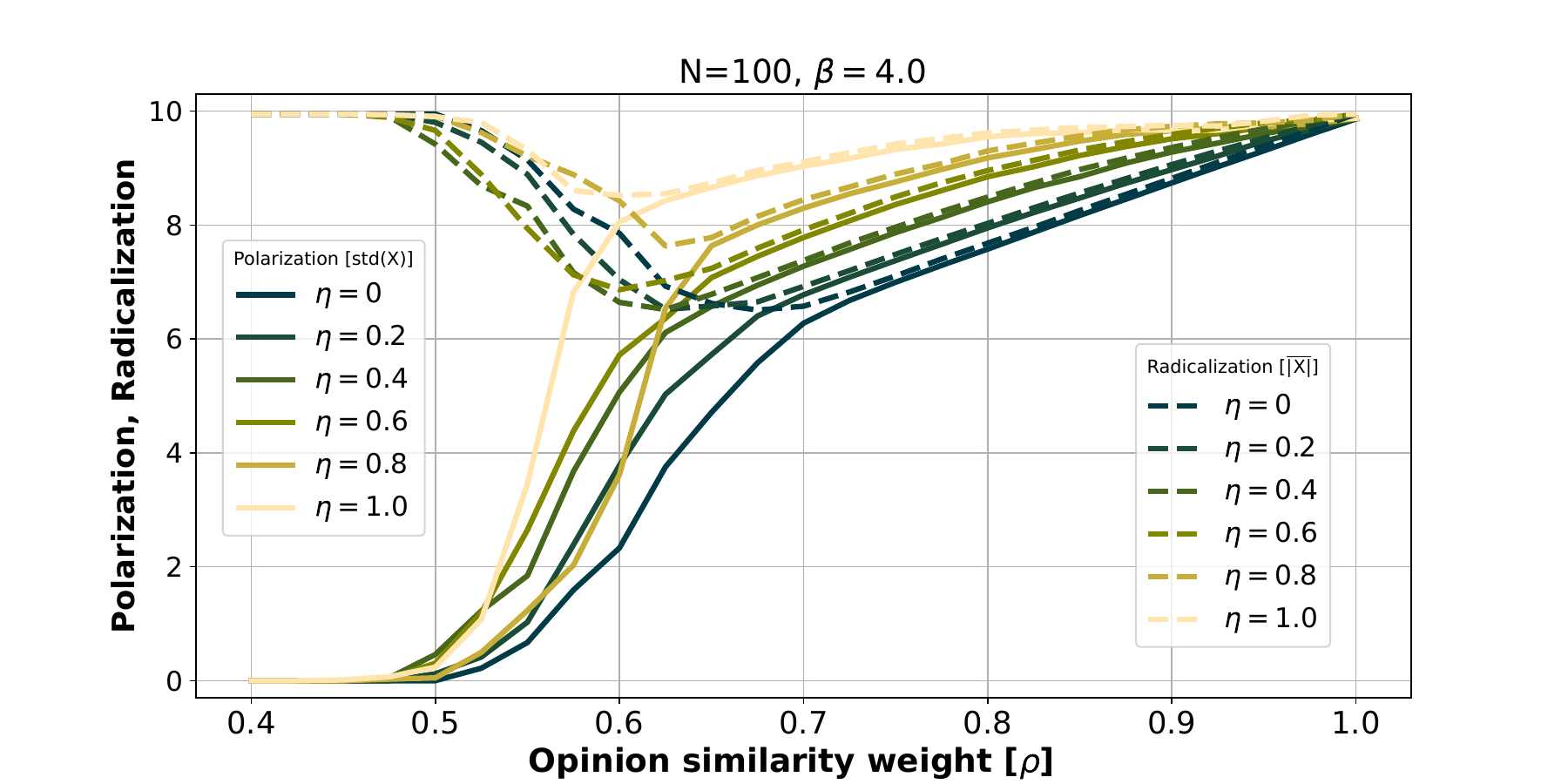}
    
\vspace{0.4cm}
    \includegraphics[width=0.25\textwidth]{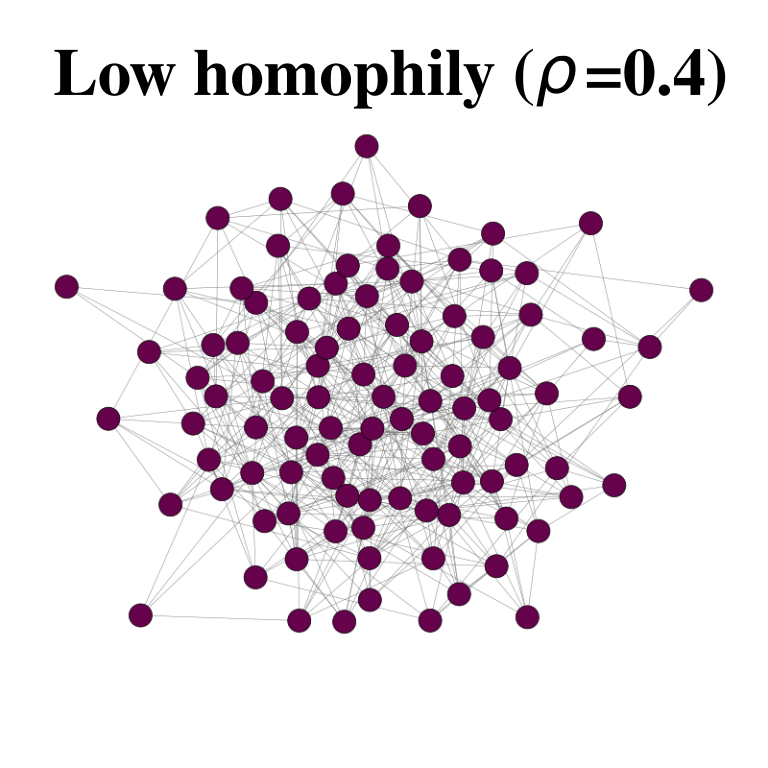}
    ~
    \includegraphics[width=0.27\textwidth]{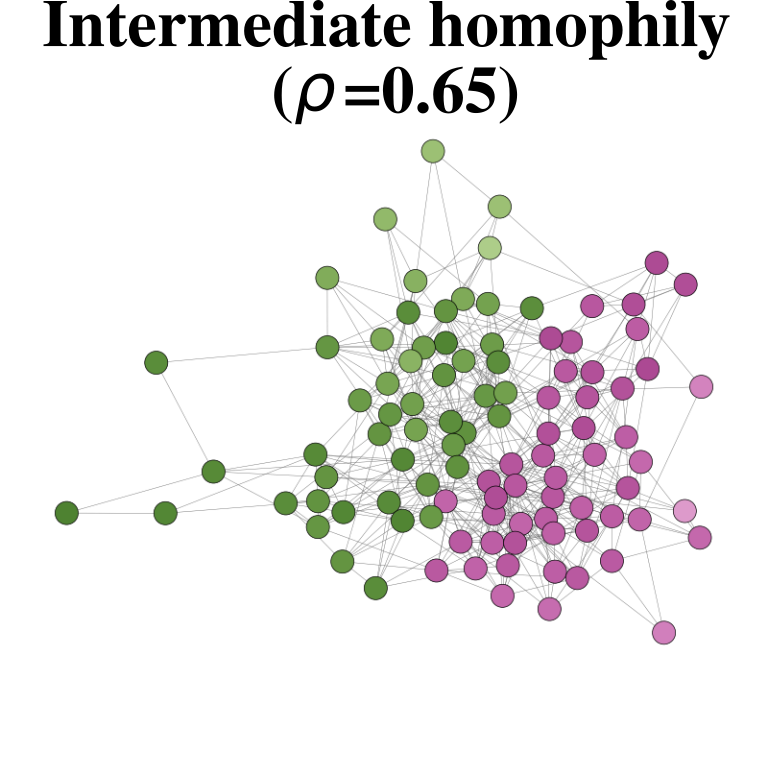}
    ~
    \includegraphics[width=0.27\textwidth]{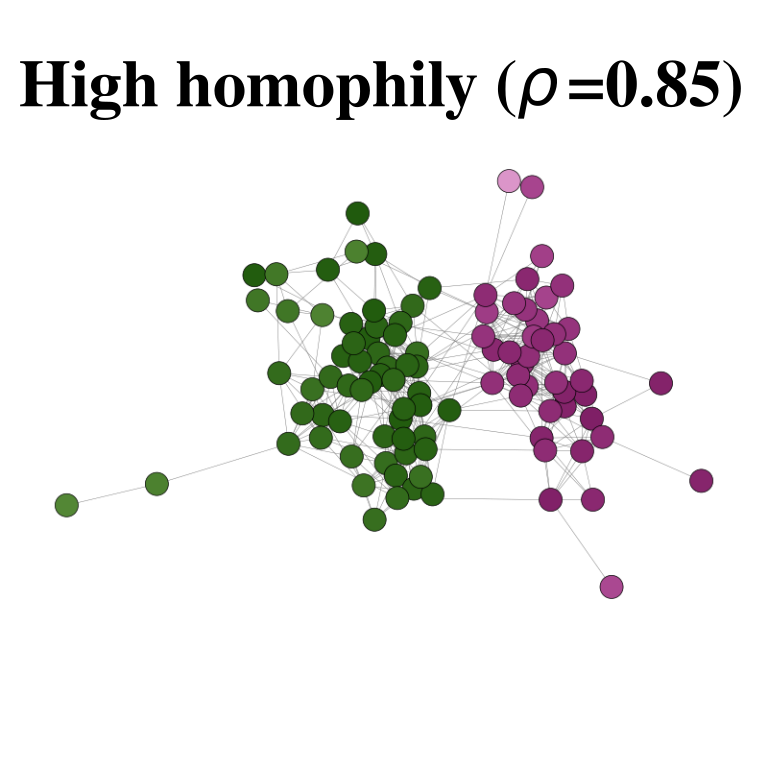}
    \caption{(Top) Polarization (solid lines) and radicalization (dashed lines) values for $\rho>0.4$. While polarization rises for all 6 values of $\eta$ shown, sublinear values are more efficient in dampening extremism, with lowest radicalization values found for $\eta<0.4$. As seen by the convex aspect of the radicalization curves, the minimum appears at intermediate values of $\rho$ that depend on other parameters. (Bottom) Network representations illustrating three different regimes observed. For low values of $\rho$ the consensus state is reached. For intermediate levels, we observe a global minimum on the radicalization curve, accounting for less extremists on the connected population. For high values of $\rho$ we observe an increase in radicalization, led by an increasing separation between groups.}
    \label{fig:SublinearStructSimi}
\end{figure}

To answer this question we allow both similarity strengths $\eta$ and $\beta$ to vary independently and focus on varying $\eta$ under high values of $\beta$. 
In Fig. \ref{fig:SublinearStructSimi}, we note that the level of opinion similarity ($\rho$) -- together with a sublinear value for structural strength ($\eta<1$) -- determines the regime in which different opinions coexist. For low values of homophily ($\rho<0.5$), we observe dominance of one single stance over the other, and the population reaches consensus. For intermediate values of homophily ($0.5<\rho<0.75$), we observe a minimum value for radicalization for that set of parameters. The corresponding network does not clearly favor one of the opinions over the other, as can be seen by two similarly sized modules with strong internal connectivity. On the other hand, for high values of opinion similarity ($\rho>0.75$), radicalization values start to rise, together with polarization, as we see individual opinions becoming more extreme (darker colors) and the modules becoming less connected to each other. We show in the Supplementary Information (also for $\beta=4$) that useful recommendations through triadic closure (positive $\eta$) are only able to lower radicalization to a minimum until a certain weight of homophily on the recommender.

As briefly discussed above, we see that strong homophily ($\rho\approx 1$) leads to fragmentation under linear similarity strength ($\eta=1$). On the other hand, favoring transitivity ($\rho<0.5$) allows for a single connected component until supra-linear conditions ($\eta>1$), resulting in a system that is more resilient against fragmentation. Together with our observations from Fig. \ref{fig:SublinearStructSimi}, this indicates that a certain level of transitivity is needed to prevent the population from breaking apart, while possibly keeping coexistence of opinions.

Although the results we discuss here are numerical and cover a limited range of network sizes, in the Supplementary Information we provide a robustness check for more parameters and an analytic study of the interplay between opinion and structural similarity metrics in preventing network fragmentation.  
Such analysis qualitatively confirms that, especially on homophilic settings, a certain level of transitivity is needed to counteract the tendency for group-wise fragmentation (see Supplementary Information, Appendix B).

\section{Conclusion}
Here we present a new model that considers both structural similarity (based on triadic closure) and opinion similarity (based on homophily) when recommending and adding links in online social networks. We aim at studying how these two distinct principles of users' similarity, when combined in link-recommendation algorithms, affect opinion dynamics and network structure.
While homophily and triadic closure are well-known mechanisms for creating social ties, their combined effects remain largely unexplored, particularly in settings where users' preferences co-evolve with network structure. Previous work \citep{abebe2022effect,bachmann2025network,asikainen2020cumulative} investigated the impact of triadic closure on network formation models that consider homophily. Under these static-feature assumptions, observations are not straightforward. While \citep{abebe2022effect, bachmann2025network} have shown that triadic closure decreases network segregation, hinting that the two principles might not work in tandem, \citep{asikainen2020cumulative} observes an increase in observed homophily.

Our results show that, despite inherent differences in the similarity metrics, both can independently lead to polarized communities when link-recommendation algorithms strongly rely on them as heuristics to suggest links. We also show that polarization characteristics differ among similarity metrics: while recommendations based on opinion similarity lead to higher polarization, rewiring based on structural similarity leads to a larger number of (dis)connected components. Nevertheless, we demonstrate that, under strong homophilic scenarios, rooting recommendations on structural similarity can provide a route to offer useful recommendations to users, while dampening extremism and enabling opinion coexistence. The above observations agree with \cite{abebe2022effect}, as they highlight that a weak but existent relationship with transitivity appears to be crucial for avoiding network fragmentation. Additionally, a recent work by \cite{bachmann2025network} argues that when triadic closure is conditional on the mechanisms of homophily and preferential attachment it exacerbates degree inequality while reducing segregation. Taken together, these results suggest that, although polarization emerges as a robust outcome of similarity-driven recommendation, incorporating structural similarity as a principle for recommendations can counteract the high polarization and radicalization outcomes resulting from high homophily.

Our findings are conditioned on two modeling choices: an opinion dynamics model that only results in polarization when due to the recommender, and simplifying assumptions of the network (fixed size, density and initial topology). Exploring other classes of opinion dynamics on our model would be a natural pathway for testing robustness of rewiring-driven polarization. For example, bounded confidence models could identify tolerance thresholds that alter the observed stable states; or anti-conformist individuals could act opposed to the averaging rule, changing the absorbing states of the dynamics (as in \cite{santos2021link} and \cite{mittal2024anticonformists}).

More broadly, our results show that design choices on online social platforms' recommender systems can deeply affect polarization and network cohesion in the long-term. When models are universally applied at a given platform, design and parameter choices can have real large-scale impacts. Here, we show that, counter-intuitively, transitivity appears to compensate for homophilic tendencies. This observation by itself incentivizes further research to investigate the matter, and leads a new way for online recommenders to be understood and designed for social media platforms.

\section*{Declarations}

\subsection*{Availability of data and materials}
Data sharing is not applicable to this article as no datasets were generated or analyzed during the current study.

\subsection*{Competing interests}
The authors have no conflicts of interest to declare that are relevant to the content of this article.

\subsection*{Funding}
This work is supported by ERC grant (RE-LINK, https://doi.org/10.3030/101116987). Views and opinions expressed are however those of the author(s) only and do not necessarily reflect those of the European Union or the European Research Council Executive Agency. Neither the European Union nor the granting authority can be held responsible for them.

\subsection*{Authors' contributions}
G.D.F. and F.P.S. contributed to the conceptualization and research design of this work. G.D.F. was involved in coding, implementing the experiments, analyzing the results and writing. F.P.S., M.C.C. and V.V.V. were involved in supervision and revision of the final draft. 
All authors read and approved the final manuscript.


\bibliography{references}

\end{document}